\documentclass[sigconf,nonacm]{acmart}

\usepackage{booktabs}
\usepackage{tikz}
\usepackage{pgfplots}
\pgfplotsset{compat=1.18}
\usepgfplotslibrary{groupplots}
\usetikzlibrary{arrows.meta,positioning,fit,backgrounds}

\AtBeginDocument{}

\begin{document}

\title{When Does Order Flow Matter? State-Dependent L2 Liquidity-State Transitions in Crypto Futures}

\author{Joohyoung Jeon}
\affiliation{%
  \institution{Korea University}
  \city{Seoul}
  \country{South Korea}
}
\email{joohyoung@korea.ac.kr}
\renewcommand{\shortauthors}{Jeon}

\begin{abstract}
Building event-conditioned market models requires separating macro-event labels from persistent microstructure state. We study this distinction in Binance BTCUSDT and ETHUSDT futures from 2023--2026, combining top-20 L2 order book data, trade-flow records, and macro-event windows. We define a supervised discrete L2 liquidity-state transition task, distinct from latent-regime detection and from price-direction prediction, and evaluate models in rolling monthly out-of-sample folds with event-clustered validation and blocked permutation tests, admitting each feature layer only if it improves on the layer below it on the same panel. Within these event windows, the first-order predictive signal is the pre-event L2 liquidity state: a coarse pre-event state baseline strongly predicts post-event liquidity regimes, interpretable logit models over continuous L2 features fail to improve on it, and a shallow nonlinear L2 model adds a robust further predictive gain of comparable size to the state baseline's own. The macro-event calendar enters only by locating the windows and supplying matched non-event controls; we use the event timing but not the event's label content as a competing predictor, so the comparison is between pre-event state and an uninformed within-window baseline, not against the event type. Order flow provides further incremental value only when layered on top of the L2 state model, not as a replacement. This value is not robustly cross-symbol: for ETH it is present across calm, mixed, and stressed regimes and largest under stressed pre-event liquidity, whereas BTC shows only isolated five-minute passes and no regime that clears at both horizons. These findings motivate a state-first design principle for market microstructure models. We provide a liquidity-state transition baseline and evaluation protocol that reinforcement-learning, execution-policy, or LLM-based context layers should be required to exceed before their added value is credited.
\end{abstract}



\maketitle


\begin{table}[t]
\centering
\caption{Model comparison. The 1m and 5m columns are out-of-sample joint improvements (Section~\ref{sec:eval}); a positive value means the model improves on its baseline. Upper panel: staged baselines on the held-out event windows, the first row coarse state vs.\ marginal, later rows vs.\ the coarse state. Lower panel: order-flow overlay (augmented vs.\ L2-only) against the flow-shuffle null.}
\label{tab:baseline}
\label{tab:overlay}
\small
\setlength{\tabcolsep}{4pt}
\resizebox{\columnwidth}{!}{%
\begin{tabular}{@{}lrrl@{}}
\toprule
Comparison & 1m & 5m & Null 95th pct.\ / interval \\
\midrule
\multicolumn{4}{@{}l}{\emph{Baselines vs.\ coarse pre-event state}}\\
Coarse state (vs.\ marginal) & $+0.034$ & $+0.045$ & --- \\
Multinomial logit & $-0.048$ & $-0.034$ & below zero \\
Ordered logit & $-0.052$ & $-0.047$ & below zero \\
Nonlinear L2-shape & $+0.044$ & $+0.060$ & $[.041,.046]$; $[.057,.062]$ \\
\midrule
\multicolumn{4}{@{}l}{\emph{Order-flow overlay vs.\ flow-shuffle null}}\\
Pooled & $+0.010$ & $+0.010$ & $+0.004$ / $+0.003$ \\
BTC (not established) & $+0.001$ & $+0.003$ & $+0.002$ / $+0.002$ \\
ETH (clears null) & $+0.020$ & $+0.016$ & $+0.006$ / $+0.003$ \\
\bottomrule
\end{tabular}}
\par\smallskip
{\footnotesize The two proper scores agree in sign in every cell; for the baseline panel the $\Delta$NLL and $\Delta$Brier components are coarse state $+0.046$/$+0.061$ and $+0.022$/$+0.028$, multinomial logit $-0.063$/$-0.044$ and $-0.033$/$-0.024$, ordered logit $-0.069$/$-0.067$ and $-0.034$/$-0.026$, and nonlinear L2-shape $+0.061$/$+0.083$ and $+0.028$/$+0.037$ at the two horizons. Both logit intervals lie entirely below zero. Throughout, point estimates and $\Delta$NLL/$\Delta$Brier components are on the held-out event windows while the bootstrap intervals and the overlay null are the full-panel cluster-bootstrap; this scope difference moves the nonlinear point by at most $0.001$ and leaves every verdict (each interval excluding zero; ETH clears, BTC not established) unchanged. The ETH increment has a ninety percent event-cluster interval excluding zero ($[.020,.023]$ at 1m, $[.015,.018]$ at 5m).}
\end{table}

\section{Introduction}
\label{sec:intro}

Event-conditioned models of financial markets are appealing because market dynamics often change around scheduled public-information releases, and a model that conditions on those moments promises sharper short-horizon predictions. Yet conditioning on an event label is not, by itself, enough. A large literature documents that scheduled macroeconomic announcements move high-frequency market quality by widening bid-ask spreads, withdrawing depth, and reshaping order flow around the release~\cite{flemingRemolona1999,balduzziEltonGreen2001,andersenBollerslevDieboldVega2003}. These are descriptive event-study effects, rather than direct predictions of the market's subsequent microstructure state. The risk for an event-conditioned predictor is that the persistent microstructure state already present \emph{before} the event dominates the short-horizon dynamics. We therefore treat scheduled-event windows not as causal instruments but as structured sampling contexts in which to study a prediction problem, and within those windows we ask whether the pre-event state carries the signal; we use the event timing to define and match the windows but do not enter the event's label content as a competing predictor.

The prediction problem we pose differs from most machine-learning work on the limit order book (LOB). That lineage, from handcrafted-feature classifiers~\cite{kerchevalZhang2015} and the FI-2010 benchmark~\cite{ntakaris2018} through deep architectures~\cite{zhangZohrenRoberts2018} and recent forecasting frameworks~\cite{briolaBartolucciAste2024,sirignanoCont2018}, almost always predicts a \emph{price} outcome. Our dependent variable is not price but the post-event \emph{liquidity state} of the book, a discrete calm, mixed, or stressed regime built from relative spread, top-20 depth, and top-20 imbalance (Section~\ref{sec:data}, Figure~\ref{fig:task}), so these models are a different line of work rather than baselines we compete against. Liquidity regimes themselves are not new, appearing as latent states, spread-deviation durations, and early-warning triggers~\cite{panayiPeters2014,hiremathHiremath2026}; the nearest is unsupervised and evaluated mainly in simulation on a single asset. Our target is instead a supervised one-step pre-event-to-post-event transition over explicit spread, depth, and imbalance terciles, evaluated out of sample across two assets and rolling months (Section~\ref{sec:related}).

Posing the target this way lets us ask a layered question rather than report a single model's score. Each candidate input, including coarse pre-event state, richer continuous L2 features, nonlinear L2 shape, and local order flow, should be admitted only if it adds predictive value \emph{after} the previous layer is already in place. We make this explicit with a staged evaluation in which every layer is compared against the layer below it on the same panel, under rolling-month out-of-sample folds, event-clustered resampling, and blocked permutation tests (Equation~\ref{eq:sequence}). This discipline matters because order flow is known to be informative for price formation~\cite{kyle1985,contKukanovStoikov2014}, and generative point-process models show that its self-excitation is itself state-dependent and stronger in disequilibrium book states~\cite{morariuPatrichiPakkanen2021}. Those results concern price impact and tick-level event intensities, whereas our question is whether order flow predicts a discrete liquidity-state transition. We therefore ask not whether order flow matters in general, but whether it still adds incremental value for liquidity-state transitions once an L2-state-and-shape baseline has been established, and whether any such value is uniform across assets and regimes.

The answers support a state-first reading. The pre-event L2 liquidity state is the primary predictive object: a coarse pre-event state baseline strongly predicts the post-event regime, an interpretable continuous-feature logit does not improve over it, and a shallow nonlinear L2-shape model adds a robust further predictive gain of comparable size to the state baseline's own. Local order flow adds further value only as an overlay on top of the L2-shape model, never as a replacement, and that value depends on both liquidity state and asset rather than being a universal signal: it is ETH-dominant and stress-amplified, while for BTC the corresponding tests do not establish an increment (Figure~\ref{fig:regime}).

\paragraph{Contributions.}
\begin{enumerate}
\item We define an event-window L2 liquidity-state transition task for BTCUSDT and ETHUSDT futures, with the target a discrete calm, mixed, or stressed regime from spread, depth, and imbalance rather than a price-direction label.
\item We evaluate a staged sequence on a common panel, from a coarse pre-event state baseline through continuous L2 features, nonlinear L2 shape, and an order-flow overlay, under rolling-month out-of-sample folds, event-clustered resampling, and blocked permutation tests, and establish that the pre-event L2 state is the first-order predictor within these event windows: continuous-feature logits do not improve over the coarse state baseline, while a shallow nonlinear L2-shape model adds a robust further predictive gain of comparable size to the state baseline's own.
\item We show that order-flow additivity is ETH-dominant and amplified under pre-event liquidity stress, while for BTC no regime clears the corresponding order-flow tests at both horizons, so the order-flow contribution is state-dependent and asset-dependent rather than robustly cross-symbol. This within-asset, regime-conditional test of whether order flow adds value once an established L2 baseline is in place is the least-occupied of the positions the three lines of prior work leave open (Section~\ref{sec:related}).
\item We provide a baseline and an out-of-sample validation-and-calibration protocol that future reinforcement-learning, execution-policy, or LLM-based context layers must exceed before their added value can be credited.
\end{enumerate}

\section{Related Work}
\label{sec:related}

We organize prior work into three lines of research that our task touches. These are machine-learning prediction on the limit order book, the predictive role of order flow and its state dependence, and microstructure studies around scheduled macroeconomic events. For each line we state what the closest work establishes and where our task remains distinct.

\subsection{Limit-order-book prediction and the liquidity-state target}

Machine-learning forecasting on the limit order book is a mature line of work, from handcrafted-feature classifiers~\cite{kerchevalZhang2015} and the FI-2010 benchmark~\cite{ntakaris2018} to deep architectures such as DeepLOB~\cite{zhangZohrenRoberts2018}, large-scale price-formation models~\cite{sirignanoCont2018}, and recent forecasting frameworks~\cite{briolaBartolucciAste2024}. Across this lineage the prediction target is almost always a price outcome, whereas our dependent variable is a discrete post-event liquidity state from relative spread, top-20 depth, and top-20 imbalance. The nearest supervised liquidity-target work forecasts a continuous liquidity-withdrawal index on equity order data under walk-forward validation~\cite{wang2025liquidity}; it shares the linear-versus-nonlinear out-of-sample comparison but not the discrete state target, the crypto setting, or the event windows.

Liquidity regimes themselves are not new. Survival models of bid-ask spread-deviation durations~\cite{panayiPeters2014} treat liquidity as a regime, and recent work detects latent microstructure regimes in the order book directly. The closest such work~\cite{hiremathHiremath2026} defines a discrete three-regime liquidity state --- stable, latent build-up, and stress --- and detects transitions into stress with positive lead time, but as a latent, generative regime construct evaluated primarily in simulation, with a preliminary application to roughly one week of limit-order-book data for a single Bitcoin pair. We differentiate from it on four explicit axes: our target is a supervised pre-event-to-post-event classification rather than latent unsupervised early detection; our liquidity state is built from explicit spread, depth, and imbalance terciles rather than an inferred latent regime; we evaluate in rolling-month out-of-sample folds with event-clustered resampling rather than a short preliminary window backed mainly by simulation; and we study two assets under scheduled macro-event windows rather than a single Bitcoin pair with no macro conditioning. A further contrast that motivates our order-flow analysis is that in their detector depth erosion, not trade flow, drives the early triggers, whereas we test directly whether order flow adds value once the L2 state is accounted for. We therefore make no claim to introduce liquidity-regime modeling.

\subsection{Order flow and its state dependence}

That order flow carries predictive information for price is a long-standing result, from the theory of informed trading~\cite{kyle1985} to the order-flow-imbalance account of short-horizon price impact~\cite{contKukanovStoikov2014} and its multi-level extension~\cite{xuGouldHowison2019}; we treat this as established prior knowledge. A separate line shows that the effect of order flow is state-dependent. Generative state-dependent point-process models couple order-flow excitation to the book state and find it stronger in disequilibrium states~\cite{morariuPatrichiPakkanen2021}. These generative intensity models, estimated on order-by-order message data, are among the nearest conceptual neighbours to our finding that ETH order flow matters more under stressed pre-event liquidity, but they characterize how order-flow intensity is generated rather than testing whether order flow still improves prediction once a liquidity-state baseline is already in place. Regime-dependent forecasting power of order-flow imbalance has likewise been documented in index futures through a stochastic price-response model~\cite{huZhang2025}, which supports state dependence but remains a continuous price-dynamics account, whereas we test whether order flow adds value for a liquidity-state target only after an order-book baseline is established.

The closest crypto-futures work in setting, rather than in method, engineers order-book and trade-flow features on Binance futures and reports that the predictive importance of these features is stable across several crypto assets~\cite{bieganowskiSlepaczuk2026}. That result targets short-horizon return and trading signals, whereas our target is the incremental contribution of order flow to a liquidity-state transition once an order-book baseline is in place. Because the two studies measure different quantities, the two findings are not in conflict: a pattern stable for one prediction target need not be stable for another. Agent-based detection of collective liquidity-taking behaviour~\cite{balcauSanchezBetancourtSarkadiVentre2024} likewise targets participant identification on simulated flow rather than prediction of the book's post-event liquidity state.

\subsection{Macro-event windows and microstructure}

A large literature documents that scheduled macroeconomic announcements affect high-frequency market quality, widening spreads, withdrawing depth, and reshaping order flow around the release~\cite{flemingRemolona1999,balduzziEltonGreen2001,andersenBollerslevDieboldVega2003}, including joint order-flow and macro-news effects on Treasury yields~\cite{pasquarielloVega2007}. We use this literature only to justify scheduled-event windows as contexts in which liquidity is likely to be stressed, and make no causal claim that macro events drive liquidity-state transitions. Even the nearest predictive, macro-conditioned order-book work~\cite{jiangLoVerdelhan2011} models jumps and price discovery in the Treasury market, not a liquidity-state transition. Our task is the specific combination none of them occupies: a supervised, out-of-sample post-event L2 liquidity-state transition, conditioned on a scheduled macro calendar. Taken together, the three lines leave an unoccupied position whose least-occupied point is a within-asset, regime-conditional test of whether order flow adds value once an established L2 baseline is in place, which is the position this paper takes.

\section{Data and Prediction Task}
\label{sec:data}

\begin{figure}[t]
\centering
\begin{tikzpicture}[
  node distance=2.2mm,
  box/.style={draw, rounded corners, align=center, inner sep=2.5pt,
              font=\footnotesize, minimum width=3.6cm},
  lab/.style={font=\scriptsize\itshape, text=black!60},
  arr/.style={-{Latex[length=1.6mm]}, semithick}]
\node[box] (win) {macro-event window};
\node[box, below=of win] (pre) {pre-event L2 state};
\node[lab, right=1.5mm of pre] {spread / depth-20 / imb-20};
\node[box, below=of pre] (post) {post-event liquidity regime};
\node[lab, right=1.5mm of post] {calm / mixed / stressed};
\draw[arr] (win) -- (pre);
\draw[arr] (pre) -- (post);
\end{tikzpicture}
\caption{We predict the post-event L2 liquidity-state transition, not price direction; both states are built from spread, depth, and imbalance with train-fold-only tercile thresholds.}
\Description{A vertical flow from a macro-event window to a pre-event L2 state (spread, depth-20, imbalance-20) to a post-event liquidity regime labelled calm, mixed, or stressed.}
\label{fig:task}
\end{figure}
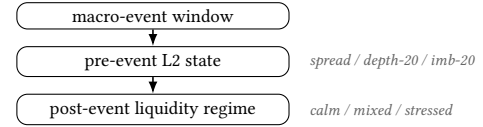

\subsection{Data and assets}

We study two perpetual-futures contracts on Binance, BTCUSDT and ETHUSDT, from January 2023 through mid-2026, using three aligned data sources: a top-20 limit-order-book record sampled once per minute, giving per-level bid and ask prices and sizes for the twenty best levels on each side; aggregate trade-flow, from which we build pre-event trade-flow summaries (signed taker volume and imbalances, trade counts, quantity, VWAP, and returns); and a calendar of scheduled macroeconomic announcements, which defines the event windows around which the task is posed. The order-book and trade-flow series are continuous over the full sample rather than restricted to event windows, and the macro calendar is used only to locate the windows, not as a model input.

The order-book record is the primary input. We normalize the twenty levels per side into a fixed-width representation and derive, at each minute, three descriptors of the book: the relative bid-ask spread, the total depth across the top twenty levels, and the order-book imbalance across those levels. These three descriptors are the basis for the liquidity-state definition in Section~\ref{sec:state}. The macro calendar is a standard commercial economic-event feed covering the major economic regions, and we use only the scheduled release timestamps. The modeling universe entering the scored results is 47{,}513 windows per horizon, 18{,}631 of them event windows (BTC 9{,}330, ETH 9{,}301) against 28{,}882 matched non-event windows, from 9{,}773 unique scheduled events across 40 monthly folds spanning February 2023 to May 2026; the pre-event state is calm, mixed, or stressed in shares of about 0.21, 0.54, and 0.25. The events are United-States-dominated (8{,}953 of the event windows) with Euro Area, Japan, the United Kingdom, and China next, and split into importance-two-or-three primary releases and importance-one controls in near-equal numbers; region and importance are carried by the calendar but are not model inputs. We make no use of the announced figures or of any surprise relative to forecast; the surprise is a potentially informative signal that we deliberately set aside, because the question is the liquidity-state transition around the scheduled event rather than the reaction to the surprise.

Our resolution is the book at one-minute cadence, sampled to twenty levels per side, chosen deliberately for two reasons. First, the venue: Binance is the most liquid venue for these contracts and so the most representative for post-event liquidity dynamics; it does distribute incremental depth feeds, but for a multi-year study at scale we use a vendor snapshot archive of the twenty best levels each minute rather than reconstructing the book from real-time updates. Second, the cadence is matched to the question: the task is a transition between liquidity states aggregated over minutes around an event, not a sub-second execution problem, and prior macro-announcement microstructure studies measure responses over windows of seconds to minutes~\cite{flemingRemolona1999,balduzziEltonGreen2001,andersenBollerslevDieboldVega2003,pasquarielloVega2007}. Being continuous over the whole sample, it also supports the event-clustered and permutation controls of Section~\ref{sec:eval} without event-window selection bias.

We are equally explicit about what this cadence does not support, so that no claim is read at a resolution it was not tested at. A one-minute snapshot cannot recover queue position, own-order fill probability, sub-second market-order impact, or a replay-grade execution simulator; those require order-by-order data that is a different research program (Section~\ref{sec:related}), and we make no such claim.

\subsection{The liquidity-state definition}
\label{sec:state}

We summarize the book at each minute by a single discrete liquidity state with three levels, which we label calm, mixed, and stressed. The state is built from the three book descriptors of Section~\ref{sec:data} (relative spread, top-twenty depth, and top-twenty imbalance) aggregated over the pre-event window. Each descriptor is first oriented so that a higher value is a less liquid book: spread and absolute imbalance are taken as is, while depth is negated, since a deep book is liquid and a thin book is stressed. A contract-minute is then assigned to a level by an equal-weight count of how many of the three oriented descriptors fall in their top tercile, capped at two so that two or three severe descriptors both map to stressed (Figure~\ref{fig:task}). The state is intentionally coarse. It is a regime label rather than a continuous score, because the question we pose is whether the regime a contract is in before an event predicts the regime it is in afterward.

Two properties of the construction matter for validity. First, the tercile thresholds are estimated per symbol from training-fold data only and applied to held-out folds, so no information from the evaluation period enters the labels, and the combined state carries more predictive content than any single descriptor (joint improvement over the marginal baseline of 0.048 at the five-minute horizon, against 0.033 for the strongest single descriptor). Second, the state is defined symmetrically before and after the event window, so the same discretization produces both the pre-event and post-event state, making the pre-to-post mapping a well-defined transition rather than a comparison of two differently constructed quantities.

\subsection{The transition task and leakage controls}

The task is to predict the post-event liquidity state from information available before the event. Each window is anchored at the release time $t$: the pre-event vector aggregates the half-open interval $[t-5\,\text{min}, t)$ and the post-event state aggregates $[t, t{+}h)$ at horizon $h \in \{1\,\text{min}, 5\,\text{min}\}$, with same-country releases within sixty seconds merged into one window. Formally, for event window $i$ and horizon $h$ the target is
\begin{equation}
\label{eq:target}
Y_{i,h} = S^{\mathrm{post}}_{i,h}, \qquad S \in \{\textsf{calm}, \textsf{mixed}, \textsf{stressed}\},
\end{equation}
where the state $S$ is derived from relative spread, top-20 depth, and top-20 imbalance using tercile thresholds estimated on the training folds only. We report two horizons, one minute and five minutes after the window, and the primary specification uses the five-minute state built from the combined descriptor. The prediction is one step and discrete, a three-class classification of the post-event regime, evaluated out of sample as described in Section~\ref{sec:eval}.

Three enforced controls keep the task honest: pre-event and post-event features are kept strictly separate, so only pre-event book information enters the predictor; all transforms and thresholds, including the tercile cut points, are fit on training folds only, ruling out global-threshold leakage across the train-test boundary; and the unit of resampling is the source event rather than the row, so minutes from the same announcement are not treated as independent.

\section{Evaluation Protocol}
\label{sec:eval}

The prediction task of Section~\ref{sec:data} admits many candidate inputs, from the coarse pre-event state through richer continuous features, nonlinear book shape, and local trade flow. The difficulty is not fitting any one of them but deciding which of them genuinely adds predictive value rather than apparent value from researcher choices. Our evaluation is built around that question. We compare a fixed sequence of models in which each layer is admitted only if it improves on the layer below it on the same data, under an out-of-sample protocol and a set of controls designed so that an added feature cannot earn credit unless its contribution survives them. This section defines the sequence, the protocol, and the controls. The results that fill them appear in Section~\ref{sec:results}.

\subsection{A staged sequence of models}

We evaluate five models in a fixed order, each adding one kind of information to the one before it (Equation~\ref{eq:sequence}). The first is a marginal baseline conditioning only on symbol and horizon class frequencies, with no order-book information. The second is a coarse pre-event state baseline conditioning only on the calm, mixed, or stressed pre-event liquidity state of Section~\ref{sec:data}; it is the empirical conditional-Markov transition model $P(S^{\mathrm{post}} \mid \mathrm{symbol}, \mathrm{horizon}, S^{\mathrm{pre}})$, fit out of sample, so it already carries pre-event-state persistence (stressed-to-stressed probability up to 0.456) and a later layer is credited only with what it adds beyond that persistence, not over an uninformed marginal. It does not condition on time of day; intraday seasonality is instead addressed in the order-flow null below. The third is a multinomial logit over the continuous L2 features, the same information in continuous form, testing whether a linear model over those features improves on the coarse state. The fourth is a depth-capped gradient-boosted classifier over multi-scale summary descriptors of the top-twenty book, the L2-shape model, which can capture nonlinear relationships among spread, depth, and imbalance that the logit cannot; it is a histogram gradient-boosted classifier kept deliberately shallow, with maximum depth three, sixty boosting iterations, learning rate 0.05, and $L_2$ regularization 1.0, so that ``shallow nonlinear'' is concrete and the comparison against the linear logit is one of model class rather than of capacity alone. The fifth adds pre-event order-flow features to the L2-shape model, giving an order-flow-augmented model whose comparison against the L2-shape model alone isolates the incremental value of order flow.

\begin{equation}
\label{eq:sequence}
\begin{aligned}
&p_{\mathrm{marg}}(Y); \quad p_{\mathrm{state}}(Y \mid S^{\mathrm{pre}}); \quad p_{\mathrm{logit}}(Y \mid X^{\mathrm{L2}}_{\mathrm{cont}});\\
&p_{\mathrm{L2}}(Y \mid X^{\mathrm{L2}}); \quad p_{\mathrm{L2{+}flow}}(Y \mid X^{\mathrm{L2}}, X^{\mathrm{flow}}).
\end{aligned}
\end{equation}

The order is deliberate: each model is the natural baseline for the next, and every comparison is made on identical observations, separating a genuine increment from a relabelling of already-captured signal.

\subsection{The out-of-sample protocol}

Every comparison is made out of sample on a common panel of held-out event windows together with matched non-event windows from the same months, shifted by a whole number of weeks so that time of day and weekday are preserved. The baseline and overlay results are reported on the event windows, and the regime decomposition of Section~\ref{sec:regime} on the full panel of event and matched non-event windows. We use rolling whole-month folds in which a model is trained on earlier months and scored on a held-out later month, so training data always precede the evaluation month. All transforms and thresholds, including the tercile cut points, are fit on the training folds only and applied unchanged to the held-out month, the leakage control of Section~\ref{sec:data} applied to every model in the sequence.

We score predictions by negative log-likelihood (NLL), with the Brier score as a secondary proper score, and summarize each comparison as the baseline NLL minus the model NLL, so a positive value means the model assigns higher likelihood to the realized post-event state (Equation~\ref{eq:score}). Because observations from the same scheduled event are not independent, we attach uncertainty with a bootstrap that resamples whole events rather than individual minutes, using the source event as the cluster unit and five hundred resamples, and we report stability across folds so a single-month-driven improvement is visible.

\begin{equation}
\label{eq:score}
\begin{aligned}
\mathrm{NLL}(M) &= -\frac{1}{N}\sum_{i} \log p_M(Y_i \mid X_i),\\
\Delta(M,B) &= \mathrm{NLL}(B) - \mathrm{NLL}(M),
\end{aligned}
\end{equation}
so that $\Delta(M,B) > 0$ means model $M$ improves on baseline $B$. The reported improvement is the joint average of this NLL improvement and the corresponding Brier improvement.

\subsection{Permutation tests and per-symbol reporting}

A positive out-of-sample improvement is necessary but not sufficient, because a flexible model can extract a small apparent gain from features that carry no genuine signal. To guard against this we add a permutation test for the layers whose value we want to establish. The test shuffles only the features under examination over sixty-four draws within blocks defined by month, symbol, and pre-event state, leaving the other features and the labels intact, and rescores the model on each shuffle, refitting it where the examined features are model inputs. This produces a distribution of improvements the same model achieves when the examined features carry no real information beyond their marginal distribution, and we require the observed improvement to exceed the upper tail of that distribution before treating it as established. For the order-flow layer in particular we shuffle only the trade-flow features and leave the order-book features in place, so that the test asks specifically whether trade flow adds information beyond the book. We also repeat this flow shuffle with the UTC hour added to the blocks, whose hourly cells keep 0.989 of held-out rows permutable, reducing the risk that order flow earns credit as an hour-of-day proxy.

We report every order-flow comparison separately for each symbol rather than only as a pooled average, and decompose the order-flow increment by pre-event liquidity regime. Pooling can let a strong result for one asset stand in for both, and a single overall number can hide that an effect is concentrated in one regime. Reporting per symbol and per regime makes any such concentration explicit, and is why the order-flow findings of Sections~\ref{sec:overlay} and~\ref{sec:regime} can be stated at the asset and regime level rather than as a single cross-symbol claim.

The combination of these elements is itself part of what this paper offers: each removes a specific way an added feature could appear useful without being so, and we present the protocol as a validation and calibration procedure for event-conditioned microstructure prediction, against which later reinforcement-learning, execution-policy, or context layers should be measured.

\section{Experiment Results}
\label{sec:results}

Sections~\ref{sec:stateshape} and~\ref{sec:overlay} report on the held-out event windows of the common panel; Section~\ref{sec:regime} reports the regime decomposition on the full held-out panel of event and matched non-event rows. Each comparison is summarized by the joint improvement of Section~\ref{sec:eval}, the mean of the negative-log-likelihood and Brier improvements over its baseline, reported at three decimals, with bootstrap intervals or flow-shuffle null thresholds where they define the test; the two proper scores agree in sign in every cell, with the separate components reported in the note to Table~\ref{tab:baseline}. The three subsections follow the staged sequence of Section~\ref{sec:eval}.

\subsection{Pre-event L2 state and nonlinear L2 shape}
\label{sec:stateshape}

The first question is whether the pre-event liquidity state predicts the post-event state at all. It does, and by a wide margin: conditioning on the coarse calm, mixed, or stressed pre-event state improves on the marginal baseline by 0.034 at the one-minute horizon and 0.045 at the five-minute horizon. The improvement holds across the panel, with the conditional state baseline beating the marginal one on roughly two thirds of the held-out rows. The pre-event liquidity state is therefore the first-order predictive object for this transition task, which is the empirical content of the state-first reading, and everything that follows is measured against this baseline rather than against the marginal one. Nor is the state baseline reducible to the train-fold-only five-minute realized-volatility tercile we test against it: that tercile improves on the marginal model by only 0.008 and 0.015 at the two horizons, while the liquidity state exceeds it by 0.026 and 0.029 (differences on unrounded values), with ninety percent event-cluster intervals above zero, so volatility is predictive here but does not reproduce the spread, depth, and imbalance state.

A natural next step is to give a linear model the same information in continuous form. We replace the coarse state with the underlying continuous L2 features and fit a multinomial logit, and an ordered-logit variant that respects the calm-to-stressed ordering. Neither improves on the coarse state baseline. The multinomial logit is worse by 0.048 at one minute and 0.034 at five minutes, and the ordered variant is worse by 0.052 and 0.047 at the two horizons, with the shortfall present for both assets at both horizons. Exposing the continuous features to a linear model does not recover the predictive value already held by the coarse state, so the continuous features help only if a model can use them nonlinearly.

A shallow nonlinear model over the same L2 features closes that gap and adds further value. A depth-capped gradient-boosted classifier over multi-scale summary descriptors of the top-twenty book --- aggregate depth and imbalance, best-quote spread, and mid-price dynamics summarized over the pre-event windows --- improves on the coarse state baseline by 0.044 at one minute and 0.060 at five minutes, with ninety percent bootstrap intervals that exclude zero (Table~\ref{tab:baseline}). Both logit intervals lie entirely below zero over the same bootstrap, so the shortfall is a stable result and not sampling noise. The nonlinear gain is present for each asset separately, improving on the coarse state baseline by 0.037 for BTC and 0.052 for ETH at one minute and by 0.052 and 0.067 at five minutes, so it is not a pooling artifact. We read this as a predictive gain in the nonlinear shape of the book, layered on the first-order state baseline rather than preceding it, even though its increment is comparable in size to the state baseline's own gain. This model classifies the three-state post-event regime with accuracy 0.586 at one minute and 0.554 at five minutes, against majority base rates of 0.571 and 0.532; with the mixed state covering about 0.55 of the rows, this is a modest but consistent edge rather than a class-balance artifact. The accuracy gain is small by construction: what improves is the calibration of the predicted state probabilities, the quantity a liquidity-aware execution or quoting system actually consumes when it weights actions by the probability of each post-event regime.

Table~\ref{tab:baseline} collects the baseline comparison. This nonlinear L2-shape model is the baseline that the order-flow overlay of Section~\ref{sec:overlay} must exceed before any order-flow value can be credited.

\subsection{The order-flow overlay}
\label{sec:overlay}

Section~\ref{sec:stateshape} fixed the bar: the L2-only nonlinear model is the baseline any order-flow contribution must clear. We add the local trade-flow features of Section~\ref{sec:eval} to the book and score the augmented model against the L2-only model on the same held-out event rows. Per symbol $s$ and horizon $h$, the increment is
\begin{equation}
\label{eq:flowincr}
\begin{aligned}
\Delta^{\mathrm{flow}}_{s,h} = \mathrm{NLL}\!\left(M^{\mathrm{L2\text{-}only}}_{s,h}\right) &- \mathrm{NLL}\!\left(M^{\mathrm{L2{+}flow}}_{s,h}\right),\\
s &\in \{\mathrm{BTC}, \mathrm{ETH}\},
\end{aligned}
\end{equation}
reported as the joint average of the NLL and Brier improvements, so it is what order flow adds once the model already sees these summary descriptors of the book.

Pooled across both assets, the overlay adds a small positive increment that clears its null. The order-flow-augmented model improves on the L2-only model by 0.010 at both the one-minute and five-minute horizons. The bar it must clear is the dedicated flow-shuffle null of Section~\ref{sec:eval}. The ninety-fifth percentile of that null is 0.004 at one minute and 0.003 at five minutes, and the pooled increment sits above it at both horizons, so order flow carries a real, if modest, increment over L2 shape pooled across the panel.

That pooled reading conceals a sharp split between the two assets, and the split is the point. For ETH the overlay is large and clears its null by a wide margin, improving on the L2-only model by 0.020 at one minute and 0.016 at five minutes, against a per-asset flow-shuffle null whose ninety-fifth percentile is 0.006 and 0.003 at the two horizons, with a ninety percent event-cluster interval that excludes zero at both horizons (Table~\ref{tab:overlay}). For BTC the same overlay is near zero and not separated from its null: the increment is 0.001 at one minute, below its null ninety-fifth percentile of 0.002, and 0.003 at five minutes against a null of 0.002, too small and horizon-specific to support a claim once the regime decomposition of Section~\ref{sec:regime} is taken into account. We therefore do not credit BTC with an established order-flow contribution: the overlay is an ETH-dominant effect, and the pooled increment is carried by ETH.

Two boundaries follow from this and are kept throughout. First, the overlay is additive, not a replacement: the order-flow-only model, given trade flow without the book, is worse than the L2-only baseline at both horizons, so order flow earns its place only as a layer on top of L2 shape. Second, the order-flow result does not transfer across assets, and we make no pooled or cross-asset order-flow claim that would average the strong ETH effect with the absent BTC one. Table~\ref{tab:overlay} reports the overlay pooled and per asset, the per-asset rows being the operative ones, and Section~\ref{sec:regime} takes the established ETH overlay apart by pre-event liquidity regime.

\subsection{Regime-conditional discovery}
\label{sec:regime}

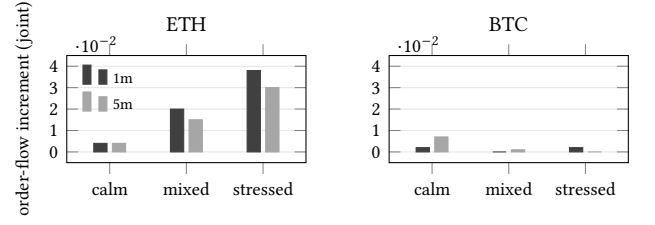
\begin{figure}[t]
\centering
\begin{tikzpicture}
\begin{groupplot}[
  group style={group size=2 by 1, horizontal sep=1.1cm},
  width=0.56\linewidth, height=3.0cm,
  ybar,
  /pgf/bar width=5pt,
  symbolic x coords={calm,mixed,stressed},
  xtick=data,
  ymin=-0.005, ymax=0.045,
  ytick={0,0.01,0.02,0.03,0.04},
  ymajorgrids, grid style={gray!20},
  tick label style={font=\footnotesize},
  label style={font=\footnotesize},
  title style={font=\small},
  enlarge x limits=0.28,
  legend style={draw=none, fill=none, font=\scriptsize,
                at={(0.03,0.97)}, anchor=north west, legend columns=1},
]
\nextgroupplot[title={ETH}, ylabel={order-flow increment (joint)}]
\addplot+[fill=black!75, draw=black!75] coordinates {(calm,0.004) (mixed,0.020) (stressed,0.038)};
\addplot+[fill=black!35, draw=black!35] coordinates {(calm,0.004) (mixed,0.015) (stressed,0.030)};
\legend{1m, 5m}
\nextgroupplot[title={BTC}]
\addplot+[fill=black!75, draw=black!75] coordinates {(calm,0.002) (mixed,-0.000) (stressed,0.002)};
\addplot+[fill=black!35, draw=black!35] coordinates {(calm,0.007) (mixed,0.001) (stressed,-0.000)};
\end{groupplot}
\end{tikzpicture}
\caption{ETH order-flow additivity is present in every pre-event regime and largest under stress; BTC is not established (only the five-minute calm cell clears its null). Bars are the joint increment over the L2-only model.}
\Description{Grouped bar chart with two panels, ETH and BTC, over calm, mixed, and stressed pre-event regimes at the one-minute and five-minute horizons. The ETH bars rise monotonically from calm to stressed at both horizons; the BTC bars stay near zero and do not rise with stress.}
\label{fig:regime}
\end{figure}

Section~\ref{sec:overlay} established the order-flow overlay as an ETH-dominant effect and left BTC uncredited. This section asks where in the range of pre-event liquidity the ETH effect lives, and whether conditioning on stress changes the BTC reading. We split the held-out evaluation rows by pre-event liquidity regime, the same calm, mixed, or stressed tercile used as the coarse state baseline, and measure the order-flow increment over the L2-only model within each regime against a flow-shuffle null computed on the same rows. For regime $r$ the within-regime increment is
\begin{equation}
\label{eq:regimeincr}
\begin{aligned}
\Delta^{\mathrm{flow}}_{s,r,h} = \mathrm{NLL}\!\left(M^{\mathrm{L2\text{-}only}}_{s,r,h}\right) &- \mathrm{NLL}\!\left(M^{\mathrm{L2{+}flow}}_{s,r,h}\right),\\
r &\in \{\textsf{calm}, \textsf{mixed}, \textsf{stressed}\},
\end{aligned}
\end{equation}
again reported as the joint average of the NLL and Brier improvements. The regime decomposition is reported over the full held-out evaluation panel, the event windows together with their matched non-event rows, rather than the event windows alone, with the real increment and its null sharing the same rows in every regime cell.

For ETH the order-flow value is present in every regime and grows with liquidity stress. The increment rises monotonically from calm to mixed to stressed, reaching 0.004, 0.020, and 0.038 at the one-minute horizon and 0.004, 0.015, and 0.030 at the five-minute horizon. Each of the six cells clears its flow-shuffle null with a ninety percent event-cluster interval above zero, and the stressed regime is the largest both in the raw increment and in its margin over the null at both horizons. The ETH order-flow overlay is therefore not a uniform additive term: it is small when liquidity is calm and several times larger when liquidity is stressed, so order flow earns its place over book shape mainly in the stressed pre-event states, which is the central discovery within the order-flow analysis.

For BTC the regime decomposition does not establish an order-flow increment. Only one of the six BTC cells clears its null, the five-minute calm regime, while the stressed regime is at or below its null at both horizons and the one-minute horizon clears in no regime. Because no BTC regime clears at both horizons, the single calm-state pass does not amount to a regime-robust effect. The borderline pooled BTC reading noted in Section~\ref{sec:overlay} resolves here: split by regime, it is carried by that lone five-minute calm cell and disappears under stress, the opposite of the stress amplification ETH shows, so the regime view is what makes the BTC non-establishment definite rather than marginal. This adjudication is unchanged when the flow shuffle adds the UTC hour to its blocks and under a Benjamini--Hochberg correction across the twelve symbol--horizon--regime cells at $q = 0.05$: every ETH cell still clears, and BTC still lacks a two-horizon pass.

Figure~\ref{fig:regime} shows the two assets side by side across the calm-to-stressed axis, and the contrast is the headline order-flow result: the contribution over the L2 state and shape baseline is asset-dependent and state-dependent, present and stress-amplified for ETH and not established for BTC, rather than a single panel-wide effect. This conditional, asset-specific reading is the empirical core of the state-first conclusion of Section~\ref{sec:discussion}.

\section{Discussion and Conclusion}
\label{sec:discussion}

The results in Section~\ref{sec:results} form a single staged, nested model comparison, and read together they point to one organizing principle for this prediction task. The central reading is a state-first one: within the event windows the calendar locates, the first-order predictor for the post-event liquidity transition is the persistent pre-event state, and anything layered on top, including order flow, should be credited only with what it adds beyond the state and shape baseline. This motivates a design principle: reinforcement-learning, execution-policy, or language-model context layers should be added only after an event-conditioned market model exceeds a robust L2 liquidity-state transition baseline of the kind established here. Two qualifications bound the reading. We use event timing to define and match windows but do not enter the event's label content -- type, region, or importance -- as a competing predictor, so the claim is that pre-event state is first-order within event windows, not that it dominates a model given the event label; testing the label attributes is future work. And the principle is shown only for the liquidity-state target; whether pre-event state is similarly first-order for a price-direction target on the same panel is untested.

The order-flow contribution should be read as conditional rather than universal: order flow adds value only when layered on the book, not as a replacement (Section~\ref{sec:overlay}), and the ETH-dominant, stress-amplified, BTC-not-established reading of Section~\ref{sec:regime} is regime- and asset-conditional rather than panel-wide. We deliberately avoid a single panel-wide statement, consistent with a literature in which order-flow informativeness is state-dependent and differs across assets.

Several boundaries should be stated plainly. This is a prediction study of liquidity-state transitions at a one-minute cadence, not a trading or execution study, and reports no policy, profit, or simulator result. The volatility-tercile, hour-blocked, and multiple-testing checks of Section~\ref{sec:results} tighten these findings without widening their scope. The two-asset scope is a genuine limit and the ETH-versus-BTC asymmetry is a two-point observation rather than a population statement; extending to additional crypto pairs, venues, and asset classes such as equities is needed to determine whether it reflects asset-specific microstructure or a broader pattern; we leave that, and sub-minute replay-grade data, to future work. The order-flow features are local trade-flow summaries rather than a full reconstruction of the process, the BTC non-establishment concerns these tests on this data rather than claiming BTC order flow is uninformative, and results are reported on the proper-score scale, not translated into basis points, profit, or any macro-event causality claim. Code, evaluation contracts, manifests, artifact hashes, and non-proprietary derived artifacts will be released upon acceptance; licensed raw inputs cannot be redistributed.

\bibliographystyle{ACM-Reference-Format}
\bibliography{refs}

\end{document}